\documentclass[preprint,showpacs,showkeywords,preprintnumbers,amsmath,amssymb,superscriptaddress]{revtex4}
\usepackage{graphicx}
\usepackage{dcolumn}
\usepackage{bm}
\begin{document}

\title{Fractional extension of Kramers rate and barrier escaping from metastable potential well}

\author{Chun-Yang Wang\footnote{Corresponding author: wchy@mail.bnu.edu.cn}}
\affiliation{Shandong Provincial Key Laboratory of Laser Polarization and Information Technology, College of Physics and Engineering, Qufu Normal University, Qufu 273165, China}
\affiliation{State Key Laboratory of Theoretical Physics, Institute of Theoretical Physics, Chinese Academy of Sciences, Beijing 100190, China}


\begin{abstract}
The reactive process of barrier escaping from the metastable potential well is studied together with the extension of Kramers' rate formula to the fractional case.
Characteristic quantities are computed for an thimbleful of insight into the near barrier escaping and recrossing dynamics.
Where the stationary transmission coefficient is revealed to be larger than the usual cases which implies less barrier recrossing.
And the non-monotonic varying of it reveals a close dependence to the fractional exponent $\alpha$.
In most cases, the near barrier behavior of the escaping dynamics is equivalent to the diffusion in the two-dimensional non-Ohmic damping system.
\end{abstract}



\maketitle

\section{INTRODUCTION}

Among all kinds of stochastic processes that produce sub-diffusion, fractional Brownian motion (fBm) \cite{fbm1} may be the model which is most particularly relevant to the study of disciplinary sciences.
For example, in the biochemistry research of subcellular transport, negative and long-range correlation has been observed in sub-diffusing mRNA molecules \cite{sd10}, RNA proteins, and chromosomal loci within E. coli cells \cite{sd04} when the fractional exponent $H$ is below $1/2$.
Meanwhile, fBms can also be used in a similar way to describe unbiased translocations \cite{sd26,sd27}, the dispersion of apoferritin proteins in crowded dextran solutions \cite{sd11}, and lipid molecules in lipid bilayers \cite{sd12}.
For these reasons, study on the model of fBm and its variances has attracted considerable attention in recent years \cite{chyx,fbm2,fbm3,fbm4,fbm5}.

However, we noticed that, in the currently existing studies of fBms, particles are always confined in a harmonically varying bounding state.
The barrier escaping of fBm particles from a quasi-bounded metastable potential well is seldom considered.
While from another point of view, barrier escape problems are of fundamental interest in physics and chemistry.
A great amount of chemical events, such as chemical reactions, molecular diffusion, or collision of molecular systems, can be modeled by a single barrier escape process within the framework of standard Brownian motion \cite{bes1,bes2,bes3}.
Great progress has been witnessed in the study of such kind of problems since the foundation of Kramers¡¯ remarkable reaction rate formula \cite{krc1}.
But no extension has been made to the fBm case.
Therefore in this paper, we report a recent study of us concerning on the fractional Kramers problems which may kill these two birds with one stone.

The paper is organized as following:
in Sec.\ref{sec2}, a briefly mathematical derivation concerning on the extension of the rate formula is presented by using of the reactive flux method \cite{rfm1,rfm2}.
In Sec.\ref{sec3}, the time-dependent transmission coefficient is computed by Laplacian solving the generalized Langevin equation of the system.
The conclusions of this study are summarized in Sec.\ref{sec4} where some prospect are also made for further discussions which may be concerned in the near future.

\section{Reactive flux formalism for the rate}\label{sec2}

For the expression of the escape rate, the traditional starting point through the method of reactive flux is alternatively to do the following integration \cite{rfm1,rfm2,jdb,jcp}
\begin{eqnarray}
k(t)=\frac{m}{Qh}\int_{-\infty}^{+\infty}dx_{0}\int_{-\infty}^{+\infty}dv_{0}W(x_{0},v_{0})v_{0}\chi(x_{0}=x_{b},v_{0};t),
\end{eqnarray}
where $Q=\int_{-\infty}^{+\infty}W_{st}(x)dx$ is the partition function of the system for reactions integrating over the distribution of the ground state $W_{st}(x)$.
$h$ denotes the phase cell and $m$ the mass of the particle.
$W(x_{0},v_{0})=\textrm{exp}[-H(x_{0},v_{0})/k_{B}T]$ is assumed to be the equilibrium distribution and $H(x_{0},v_{0})=\frac{1}{2}m(\omega_{b}^{2}x_{0}^{2}+v_{0}^{2})$
is the Hamiltonian of the system.

The function $\chi(x_{0},v_{0};t)$ is a characteristic quantity computed corresponding to an ensemble of diffusing particles which are assumed to start from $x_{0}=x_{b}$.
Mathematically it can be resulted from the following integration on the reduced distribution function
\begin{eqnarray}
\chi(x_{0},v_{0};t)=\int^{\infty}_{0}W(x,x_{0},v_{0};t)dx=\frac{1}{2}\textrm{erfc}\left(-\frac{\langle
x(t)\rangle}{\sqrt{2\sigma_{x}^{2}(t)}}\right),\label{eq,chsi}
\end{eqnarray}
where
\begin{eqnarray}
W(x,x_{0},v_{0};t) =\frac{1}{\sqrt{2\pi \sigma_{x}^{2}(t)}}\textrm{exp}\left[{-\frac{(x-\langle x(t)\rangle)^{2}}{2\sigma_{x}^{2}(t)}}\right], \label{eq,pd}
\end{eqnarray}
is reduced from the joint probability density function (PDF) of the system \cite{adelm}
\begin{eqnarray}
W(x,v,x_{0},v_{0};t)
=\frac{1}{2\pi|\mathcal{A}(t)|^{1/2}}e^{-\frac{1}{2}\left[y^{\dag}(t)
\mathcal{A}^{-1}(t)y(t)\right]},\label{eq,pdf}
\end{eqnarray}
with $y(t)$ the vector $[x-\langle x(t)\rangle,v-\langle
v(t)\rangle]$ and $\mathcal{A}(t)$ the matrix of second moments in which $\sigma^{2}_{x}(t)=\langle[x-\langle x(t)\rangle]^{2}\rangle$ is the variance of $x(t)$.

In the spirit of reactive flux calculation, the initial conditions are assumed to be at the top of the barrier.
The rate is accounted from an ensemble of reactive trajectories which start with identical initial conditions but experience different stochastic histories.
Then all the dynamics are contained in the generalized fractional Langevin equation (GFLE) \cite{fle1,fle2,fle3} of the system
\begin{eqnarray}
m\ddot{x}+\int^{t}_{0}\eta(t-t')\dot{x}(t')dt'+\partial_{x}U(x)=\xi(t)
\label{fle}
\end{eqnarray}
where $\eta(t)=\eta_{\alpha}t^{-\alpha}/\Gamma(1-\alpha)$ is the frictional kernel with fractional exponent $0<\alpha<1$ and
$\eta_{\alpha}$ the strength constant.
$\xi(t)$ is the fractional Gaussian noise with $\langle\xi(t)\xi(t')\rangle=k_{B}T\eta_{\alpha}|t-t'|^{-\alpha}$ satisfying the Kubo's fluctuation-dissipation theorem \cite{fdt1}.

Eq.(\ref{fle}) is one of the alternative form of GFLE for fBm systems which can also be derived mathematically from the system-plus-reservoir model of harmonic oscillators \cite{hotb,sdj1}.
Where $U(x)$ is the external potential which is always assumed to be a metastable form in the study of barrier escaping dynamics.
But the approximation of an inverse harmonic potential is always made in the neighbourhood of the saddle point.
Supposing the temperature of the system is much less than the barrier height of the potential, a Boltzmann form of the stationary probability distribution $W_{st}(x)$ will be satisfied in the neighbourhood of the potential well.
Then all the quantities can be obtained from Laplacian solving the GFLE.

In order to isolate the dynamical corrections to the transition state theory (TST) rate $k^{TST}=\frac{k_{B}T}{Qh}e^{-V_{_{B}}/k_{B}T}$ \cite{TST1,TST2,TST3}, it is convenient to define the transmission coefficient
\begin{eqnarray}
\kappa(t)=\frac{m}{k_{B}T}\int_{-\infty}^{+\infty}dv_{0}v_{0}e^{mv_{0}^{2}/2k_{B}T}\chi(v_{0};t),
\end{eqnarray}
Substituting Eq.(\ref{eq,chsi}) into the above integration we obtain immediately
\begin{eqnarray}
\kappa(t)=\left(1+\frac{m\sigma_{x}^{2}(t)}{k_{B}T H^{2}(t)}\right)^{-1/2},
\end{eqnarray}
where $H(t)=\mathcal{L}^{-1}[(s^{2}+s\eta(s)-\omega^{2})^{-1}]$ is namely the response function obtained from solving the GFLE and $\eta(s)=\eta_{\alpha}s^{\alpha-1}$ is the Laplacian transformation of the friction kernel $\eta(t)$.
This expression for the fractional reactive index leads immediately to Kramers¡¯ formula for the rate constant \cite{rfm1}
which physically describes the probability of a particle already escaped from the well to recross the barrier.

\section{Fractional transmission coefficient}\label{sec3}

\begin{figure}[ht]
\includegraphics[scale=0.9]{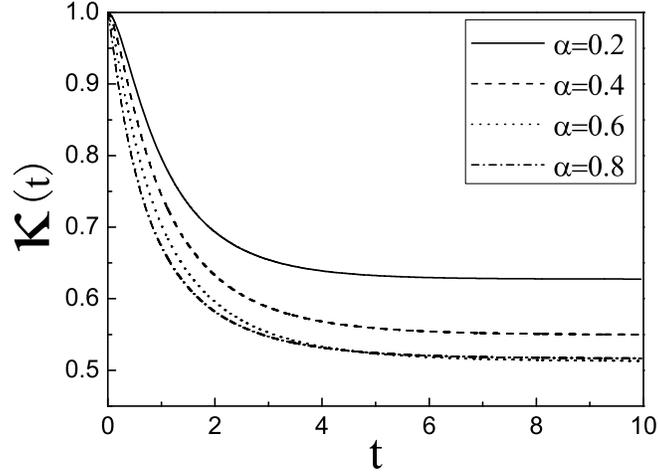}
\caption{Time dependent varying of $\kappa(t)$ at various $\alpha$. Parameters in use are
$\eta_{\alpha}=2.0$, $\omega=k_{B}T=1.0$, $x_{0}=-1.0$ and $v_{0}=2.0$. \label{fig1}}
\end{figure}

In the numerical computations here and following, all the quantities are rescaled to dimensionless ones for simplicity.
Firstly in Fig.\ref{fig1}, we plot the time-dependent transmission coefficient $\kappa(t)$ at various fractional exponent $\alpha$.
From which we can see that, the stationary value of $\kappa(t)$ (here it is defined as $\kappa_{st}=lim_{t\rightarrow\infty}\kappa(t)$) is comparable in most cases to the results of two-dimensional non-Ohmic damping systems \cite{jcp}.
However it is always larger than the usual cases such as that of one-dimensional non-Ohmic damping \cite{jdb}.
This reveals that the fBm particles already escaped from the potential well will have less probability to recross the barrier.

\begin{figure}[ht]
\includegraphics[scale=0.7]{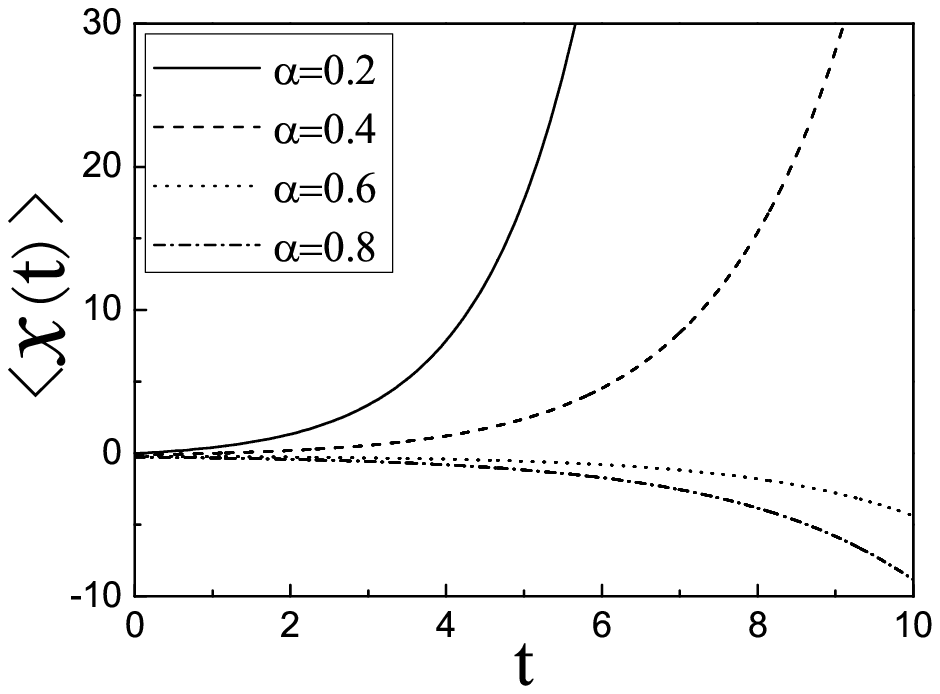}
\includegraphics[scale=0.7]{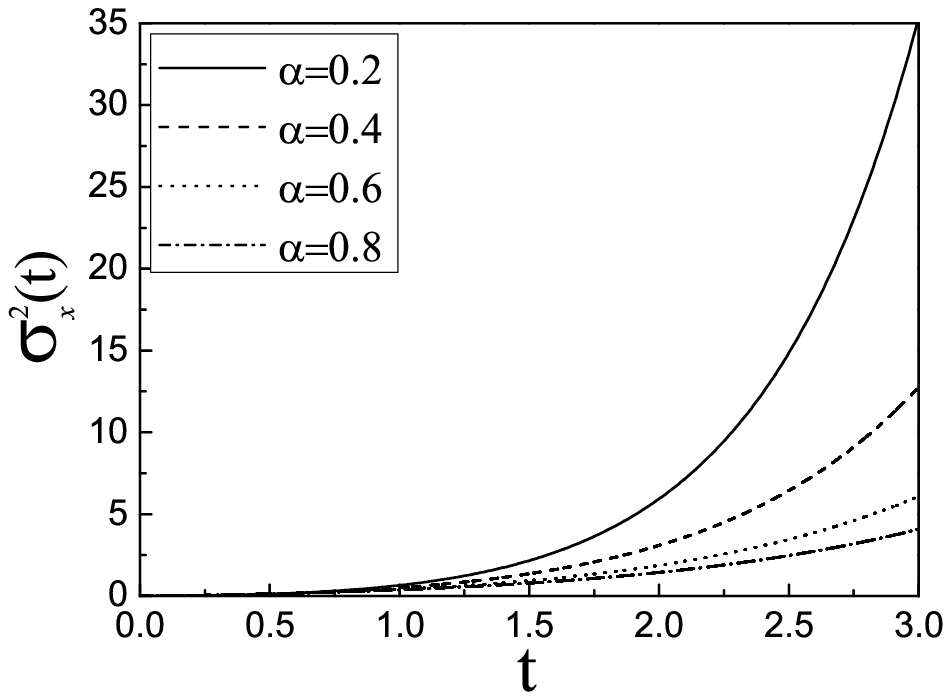}
\caption{Time dependent varying of $\langle x(t)\rangle$ and $\sigma^{2}_{x}(t)$ at various $\alpha$. \label{fig2}}
\end{figure}

In order to find the inherent reason for the occur of this physical scenario, we give an illustration on the time dependence of $\langle x(t)\rangle$ and $\sigma^{2}_{x}(t)$ in Fig.\ref{fig2}.
Seen from it that the mean displacement $\langle x(t)\rangle$ diverges negatively at $\alpha$.
However, the value of $A_{11}(t)$ is always positive increasing.
This reveals, from the view point of PDF evolution, the center of the wave packet may move in the opposite direction of diffusion.
As the forwarding of the center the width the PDF is always expanding.
Therefore, a steady barrier escaping probability can always be expected.
But the barrier recrossing is greatly reduced.
The occurrence of such a nontrivial result may probably be caused by the ``Joseph effect'' (long-range correlation) immersed in the process of fBm \cite{fbm2}.

\begin{figure}[ht]
\includegraphics[scale=0.9]{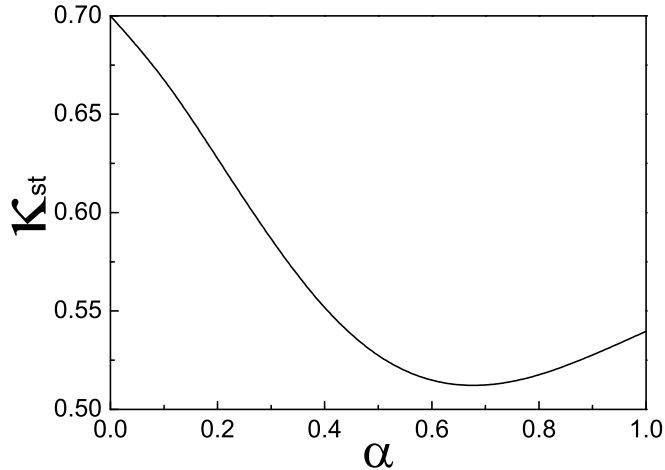}
\caption{Stationary value $P_{\textrm{st}}$ plotted as a function of $\alpha$. \label{fig3}}
\end{figure}

For more details, we plot in Fig.\ref{fig3} the $\alpha$ dependence of $\kappa_{st}$.
From which we can see that $\kappa_{st}$ varies non-monotonically as the increasing of $\alpha$.
There lives a minimum of $\kappa_{st}$ at $\alpha\simeq0.68$ in the particular case of what is considered here in this paper.
This corresponds to a maximum barrier recrossing in the process of fBm which can also be understood from considering on the reversed movement of diffusion particles mentioned hereinbefore.
Seen from Fig.\ref{fig2}, both $\langle x(t)\rangle$ and $\sigma^{2}_{x}(t)$ vary slowly as the time goes on.
This reveals that the particles will have relatively more time wandering in the neighbourhood of the saddle point.
Thus results in a maximum barrier recrossing and a relatively lower net escaping rate.

\section{summary and discussion}\label{sec4}

In conclusion, we have extended the remarkable Kramers' rate formula to the case of fractional Brownian motion and studied the near barrier dynamics of a reactive fBm process.
Where several nontrivial properties are revealed such as less barrier recrossing, a non-monotonically varying stationary transmission coefficient in great dependent to the fractional exponent $\alpha$ and so on.
The escaping of fBm particles from the metastable potential well is found to be more smoothly than the usual cases such as those in one-dimensional non-Ohmic damping systems.
And in most cases, the escaping is equivalent to the diffusion in the two-dimensional non-Ohmic damping system.
We give a brief explanation on the inherent mechanism of these results from the view point of PDF evolution.
It is expected that the present study may have provide more useful information for the in-depth understanding of anomalous diffusion dynamics and fBms.

\begin{figure}[ht]
\includegraphics[scale=0.9]{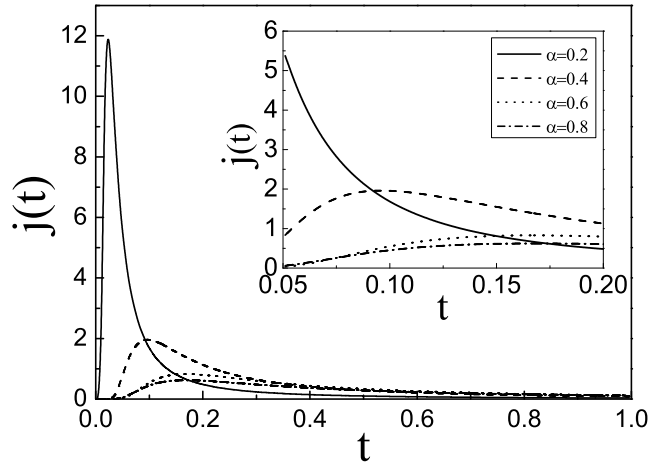}
\caption{Time-dependence of $j(t)$ plotted at various $\alpha$ in which the inserted subgraph is the local magnification of it. \label{fig4}}
\end{figure}

Yet of course, the courtship concerning on the characteristics of fBm dynamics should be more than obtaining a reactive rate.
Many facts are still left to be revealed.
For an example, one can define a reactive flux $j(t)=d\chi(x_{0}=x_{b},v_{0}, t)/dt$ near the saddle point.
And find the time-dependent varying of it for various $\alpha$ just as we have plotted in Fig.\ref{fig4}.
From which we can see that in each cases $j(t)$ approaches its maximum quickly and then vanishes to zero monotonically in the long time without any other variance.
This is reminiscent to the remarkable Kramers' rate problem and can also be understood easily from the view point of PDF evolution.

That is to say, at the beginning of the fBm diffusion process, the center of the PDF moves quickly to the forward.
And at the same time, the width of it expands in a high speed as well.
Thus results in a rapid emergence of each maximum of $j(t)$.
After a short period of time, the center of the wave packet may have passed the saddle point.
But due to the expansion of its width, the long tail of the wave packet still covers on the barrier.
Then a small flux $j(t)$ which is close to zero can persistently be found in the long time.
Along this road, any other dynamical details of the diffusion process can be found without difficulty.
Therefore we believe that there lives a great possibility for the present study to stimulate other in-depth investigations concerning on the fBms in the near future.

\section * {ACKNOWLEDGEMENTS}

This work was supported by the Shandong Province Science Foundation for Youths under the Grant No.ZR2011AQ016, the Open Project Program of State Key
Laboratory of Theoretical Physics, Institute of Theoretical Physics, Chinese Academy of Sciences, China under the Grant No.Y4KF151CJ1.


\begin{thebibliography}{References}

\bibitem{fbm1} B. Mandelbrot and J. W. van Ness, SIAM Rev. \textbf{10}, 422 (1968).

\bibitem{sd10} M. Magdziarz, A. Weron, K. Burnecki and J. Klafter, Phys. Rev. Lett. \textbf{103}, 180602 (2009).

\bibitem{sd04} S. C. Weber, A. J. Spakowitz and J. A. Theriot, Phys. Rev. Lett. \textbf{104}, 238102 (2010).

\bibitem{sd26} A. Zoia, A. Rosso and S. N. Majumdar, Phys. Rev. Lett. \textbf{102}, 120602 (2009).

\bibitem{sd27} J. L. A. Dubbeldam, V. G. Rostiashvili, A. Milchev and T. A. Vilgis, Phys. Rev. E \textbf{83}, 011802 (2011).

\bibitem{sd11} J. Szymanski and M. Weiss, Phys. Rev. Lett. \textbf{103}, 038102 (2009).

\bibitem{sd12} J.-H. Jeon, H. Martinez-Seara Monne, M. Javanainen and R. Metzler, Phys. Rev. Lett. \textbf{109}, 188103 (2012).

\bibitem{chyx} C. Y. Wang, X. M. Zong, H. Zhang and M. Yi, Phys. Rev. E \textbf{90}, 022126 (2014).

\bibitem{fbm2} I. I. Eliazar and M. F. Shlesinger, Physics Reports \textbf{527}, 101 (2013).

\bibitem{fbm3} P. Embrechts and M. Maejima, \textit{Selfsimilar Processes} (Princeton University Press, Princeton) (2002).

\bibitem{fbm4} D. Boyer, D. S. Dean, C. M. Monasterio and G. Oshanin, Phys. Rev. E \textbf{87}, 030103(R) (2013).

\bibitem{fbm5} S. Carmi and E. Barkai, Phys. Rev. E \textbf{84}, 061104 (2011).

\bibitem{bes1} P. H\"{a}nggi, P. Talkner, and M. Borkovec, Rev. Mod. Phys. \textbf{62}, 251 (1990).

\bibitem{bes2} E. Pollak, J. Chem. Phys. \textbf{85}, 865 (1986).

\bibitem{bes3} P. Talker, E. Pollak, and A. M. Berhkovskii, Chem. Phys. \textbf{235}, 1 (1998).

\bibitem{krc1} H. A. Kramers, Physica (Utrecht) \textbf{7}, 284 (1940).

\bibitem{rfm1} D. J. Tannor and D. Kohen, J. Chem. Phys. \textbf{100}, 4932 (1994).

\bibitem{rfm2} D. Kohen and D. J. Tannor, J. Chem. Phys. \textbf{103}, 6013 (1995).

\bibitem{jdb} J. D. Bao, J. Chem. Phys. \textbf{124}, 114103 (2006).

\bibitem{jcp} C. Y. Wang, J. Chem. Phys. \textbf{131}, 054504 (2009).

\bibitem{adelm} S. A. Adelman, J. Chem. Phys. \textbf{64}, 124 (1976).

\bibitem{fle1} E. Lutz, Phys. Rev. E \textbf{64}, 051106 (2001).

\bibitem{fle2} W. T. Coffey, Y. P. Kalmykov, and J. T. Waldron, \textit{The Langevin Equation}, 2nd ed. (World Scientific, Singapore, 2004).

\bibitem{fle3} S. Burov and E. Barkai, Phys. Rev. Lett. \textbf{100}, 070601 (2008).

\bibitem{fdt1} R. Kubo, Rep. Prog. Phys. \textbf{29}, 255 (1966).

\bibitem{hotb} R. Zwanzig, J. Stat. Phys. \textbf{9}, 215 (1973); R. Zwanzig, \textit{Nonequilibrium Statistical Mechanics}, (Oxford University Press, Oxford, 2001).

\bibitem{sdj1} U. Weiss, \textit{Quantum Dissipative Systems}, 2nd ed. (World Scientific, Singapore, 1999).

\bibitem{TST1} T. Seideman and W. H. Miller, J. Chem. Phys. \textbf{95}, 1768 (1991).

\bibitem{TST2} J. M. Sancho, A. H. Romero, and K. Lindenberg, J. Chem. Phys. \textbf{109}, 9888 (1998).

\bibitem{TST3} E. Pollak and M. S. Child, J. Chem. Phys. \textbf{72}, 1669 (1980).

\end{thebibliography}
\end{document}